\documentclass[a4paper]{jpconf}
\usepackage{graphicx}
\begin{document}
\title{On orbit alignment and diagnostics for the LISA Technology Package}

\author{A F Garc\'ia Mar\'in$^1$, V Wand$^1$, F Steier$^1$, F Guzm\'an Cervantes$^1$,\\
 J Bogenstahl$^{1,3}$, O Jennrich$^2$, G Heinzel$^1$ and  
K Danzmann$^{1}$} 
\address{$^1${} Max-Planck-Institut f\"ur Gravitationsphysik (Albert-Einstein-Institut) and Universit\"at Hannover, Callinstr. 38, D-30167 Hannover, Germany}
\address{$^2${} ESTEC, Noordwijk, The Netherlands}
\address{$^3${} University of Glasgow, UK}
\ead{antonio.garcia@aei.mpg.de}
\begin{abstract}
This paper presents a procedure to perform fully autonomous on orbit alignment of the interferometer on board the LISA technology package (LTP). LTP comprises two free-floating test masses as inertial sensors that additionally serve as end-mirrors of a set of interferometers.
From the output signals of the interferometers, a subset has been selected to obtain alignment information of the test masses. 
Based on these signals, an alignment procedure was developed and successfully tested on the engineering model of the optical bench. 
Furthermore, operation procedures for the characterization of critical on orbit properties of the optical metrology system (e.\,g. fiber noise) have been established.
\end{abstract}.

\section{Introduction}

The Laser Interferometer Space Antenna (LISA) is a joint ESA/NASA mission to detect and observe gravitational waves. Its technology demonstrator LISA Pathfinder carries two payloads, a European LISA technology Package (LTP) and a US-provided Disturbance Reduction System (DRS) to test key technology needed for LISA. A core part of the LTP is the Optical Metrology System (OMS) that monitors the position, the alignment and the fluctuations of the test masses~(\cite{ghh_am4}~to~\cite{Robertson_LISA}). 

The OMS includes four heterodyne interferometers on an optical bench (see Figure~\ref{layout}):
\begin{figure}[ht]
\begin{center}
\includegraphics[width=18pc,angle=90]{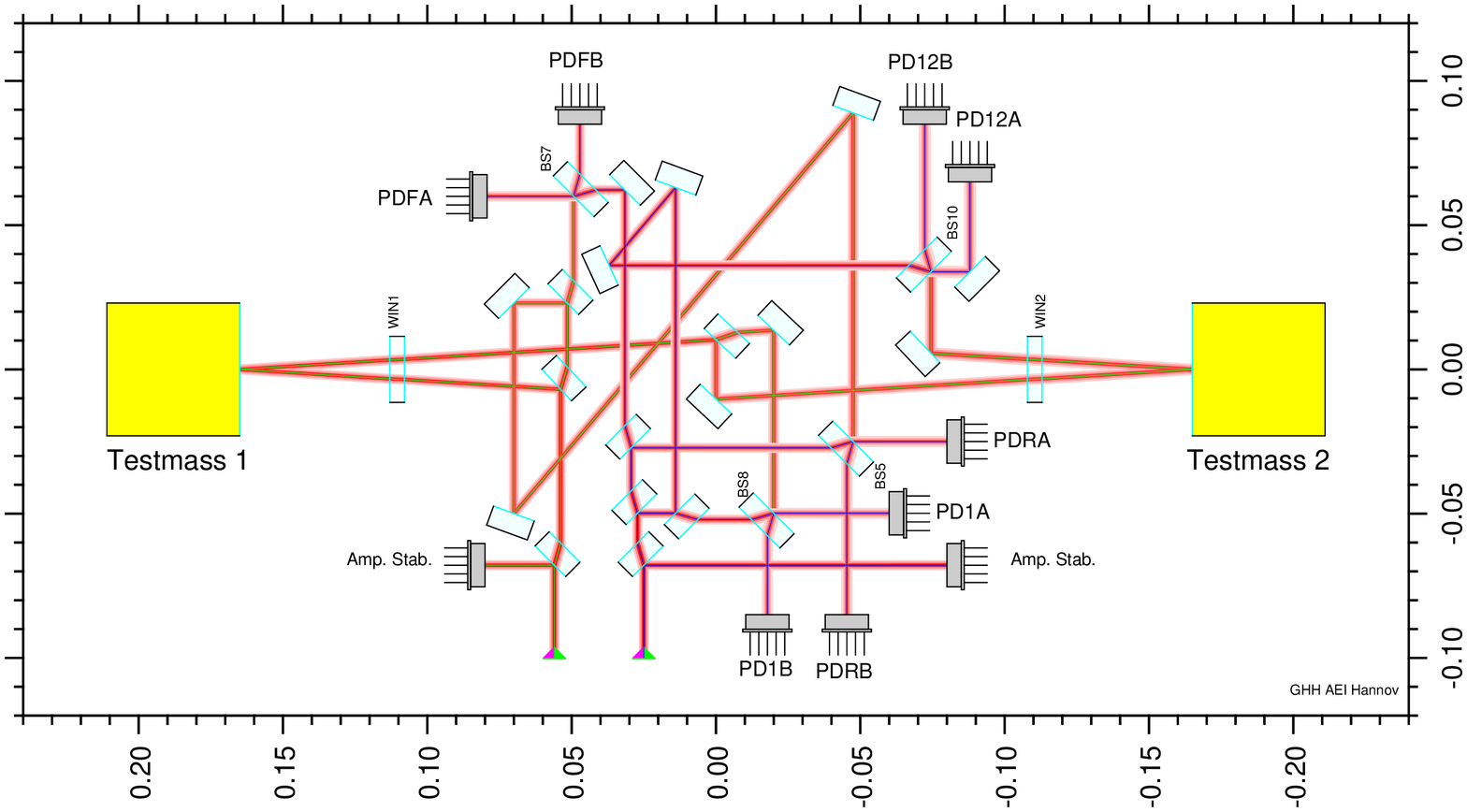}
\caption{\label{layout}Optical layout of the LTP interferometer. The squares on the sides represent the two test masses, whose position fluctuations and alignment are monitored by the set of interferometers. One photodiode is placed at each of the two output ports of each interferometer (A and B) for redundancy. \vspace{-2pc}}
\end{center}
\end{figure}
\begin{description}
\item[Reference (\textbf{R}):] This interferometer provides the phase reference for the other three interferometers.
\item[\textbf{$x_1$} (\textbf{1}):] The distance and alignment of test mass one with respect to
the optical bench.
\item[\textbf{$x_1-x_2$} (\bf{12}):] The distance between the two test masses and their mutual 
alignment.
\item[Frequency (\bf{F}):] This interferometer measures the laser frequency fluctuations
via an intentional armlength difference. 
\end{description}

\section{Alignment signals}
Quadrant photodiodes (QPD) at the output of interferometers \textbf{1} and \textbf{12} are used to obtain alignment information of the test masses. There are two independent types of alignment signals:
\begin{itemize}
\item{Position of the beam (here labelled as \textbf{DC}) with respect to the center of the QPD}
\item{Differential wavefront sensing (\textbf{DWS})~\cite{ghh_am5} that gives information on the misalignment of the beams with respect to each other, and consequently of the test mass.}
\end{itemize}
The signals selected for the alignment procedure are:

\begin{description}
\item[Total power $\Sigma$:] is the sum of the averaged power measured by each quadrant. This is a non-negative number, which is scaled such that a value of 1.0 (nominal value) indicates when both the Measurement Beam (MB) and Reference Beam (RB) are switched on and well-aligned. Experimental values from other runs or the best available prediction are used for this scaling. 
\item[Horizontal DC alignment $\phi^{\rm DC}$:] provides the difference between the averaged power measured by the left and right section of the QPD. It is normalized by the unscaled $\Sigma$, such that the variation range is $-1\dots1$ (0 at the center of the QPD).
\item[Vertical DC alignment $\eta^{\rm DC}$:] provides the difference between the averaged power measured by the upper and lower section of the QPD. It is normalized by the unscaled $\Sigma$, 
such that the variation range is $-1\dots1$ (0 at the center of the QPD).
\item[Horizontal DWS alignment $\phi^{\rm DWS}$:] gives the difference, in radian, between the phase measured by the left and right section of the QPD.
\item[Vertical DWS alignment $\eta^{\rm DWS}$:] gives the difference, in radian, between the phase measured by the upper and lower section of the QPD.
\item[Contrast on the QPD $c$:] provides the contrast measured over the whole surface of the QPD. This is a number between 0 and 1 (also usual to be expressed as 0\,\%\dots100\,\%).
\item[Longitudinal phase $\varphi$:] is the phase measured, in radians ($-\pi\dots\pi$), over the whole surface of the QPD. A phase-tracking algorithm is applied to the data, in order to avoid $2\pi$-hopping, and be able to follow long-term drifts of the test-masses.
\end{description}

Raw data delivered by the phasemeter~\cite{ghh_LISA,ghh_am6} is used to generate these signals in the Data Management Unit (DMU). These signals are labelled with a lower index `1', `12', `R' or `F' to indicate from which interferometer they originate. For example, interferometer 1 produces the signals $\Sigma_1$, $\phi^{\rm DC}_1$, $\eta^{\rm DC}_1$,
$\phi^{\rm DWS}_1$, $\eta^{\rm DWS}_1$, $c_1$ and $ \varphi_1$.

DC signals have a larger dynamic range and are not necessarily zero for optimal alignment, as they are referred to the center of the QPD. DWS signals are zero for an optimal alignment, which is to first order independent of the diode position~\cite{morrison1}, and offer better precision over a shorter dynamic range.

\section{Interferometer initial acquisition}

The aim of this procedure is to autonomously align both test masses on board LTP, by using the interferometric signals mentioned before, for initial alignment during commissioning and for later re-alignment. 

There are three different acquisition steps to be sequentially executed according to the alignment state (see Figure~\ref{diag}):
\begin{figure}[ht]
\begin{minipage}[h]{28pc}
\includegraphics[width=32pc]{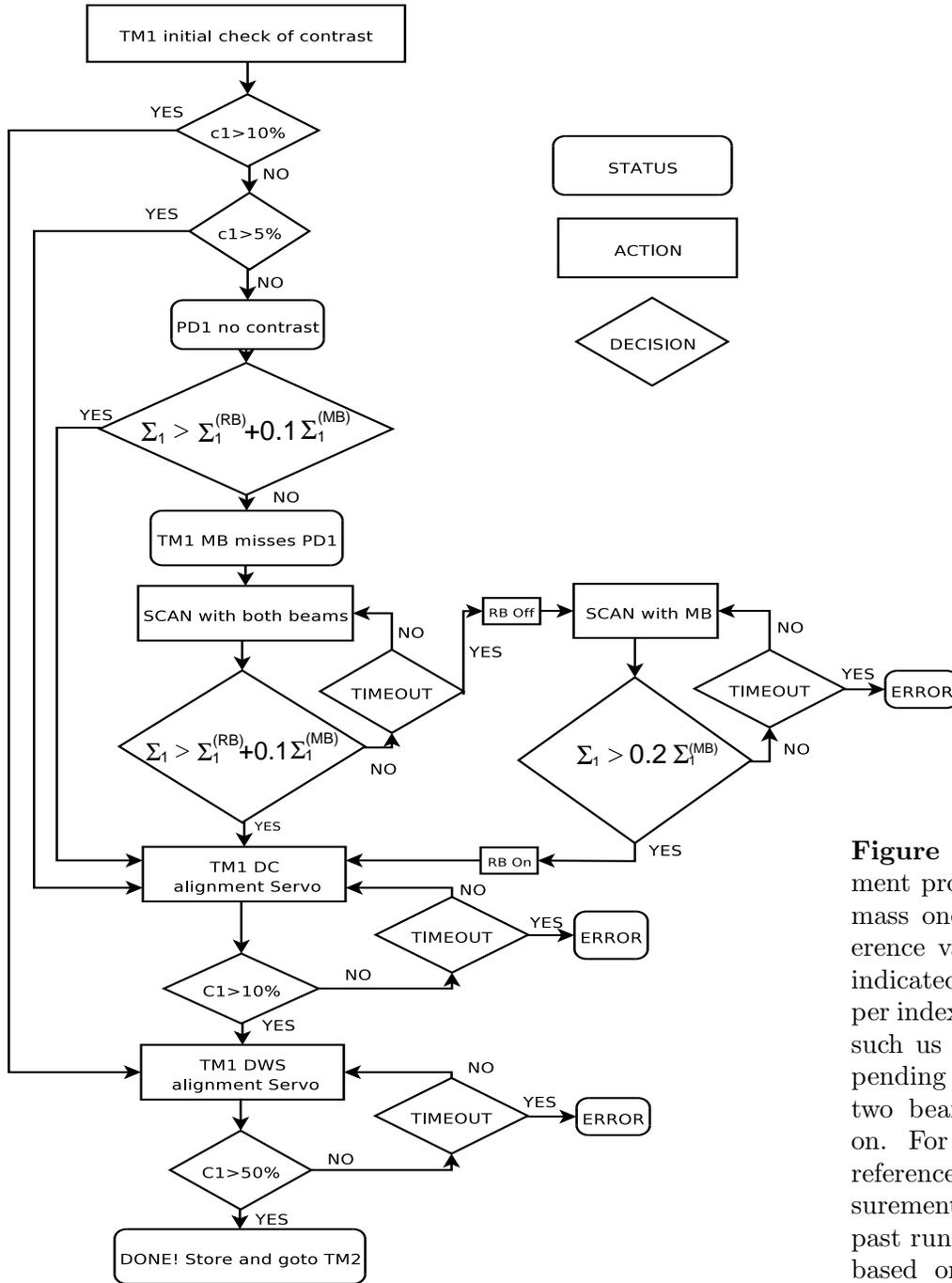}
\end{minipage}
\begin{minipage}[h]{9.5pc}\vspace{30pc}\caption{\label{diag}
Alignment procedure for test mass one (TM1). Reference values for $\Sigma$ are indicated with an upper index in parentheses such us RB or MB depending on which of the two beams is switched on. For this and other reference values, measurements stored during past runs or predictions based on ground tests are used.  
}
\end{minipage}
\end{figure}

\begin{enumerate}
\item[\textbf{Scan:}] At this step, only the reference beam hits the QPD. The test mass is moved following a spiral around the nominal incoming beam axes. This way, the reflected measurement beam describes an spiral on the detection plane until a certain percentage of it is detected by the QPD. If the process exceeds a certain pre-defined time, a time-out flag is set: the reference beam is turned off and the scanning process repeated. After a second "time-out" event, an ERROR message signalizes a malfunction of the Optical Metrology System, allowing the intervention from ground.

\item[\textbf{DC:}] The measurement beam hits the QPD, but no interference takes place. The test mass is aligned using a control loop with the DC alignment signals as error signals. The target is not to reach zero but a value pre-estimated in earlier runs or the best theoretical prediction. 
\item[\textbf{DWS:}] Both beams interfere to a level such that a predefined contrast threshold has been achieved. The test mass is aligned using a control loop with the DWS alignment signals as error signals. The target is to reach zero, as this means optimal overlap of the two interfering beams. This alignment has a higher sensitivity in comparison to the DC alignment but a smaller dynamic range.
\end{enumerate}

Each interferometer is considered to be properly aligned if it delivers contrast values greater than 50\,\% and the DWS signals reach zero.

\subsection{Experimental implementation}
The alignment procedure was tested on the Engineering Model of the LTP optical bench, for which test mass 1 was substituted by a 3-axis PZT. 
Alignment signals were produced by dedicated FPGA-based phasemeter~\cite{ghh_LISA,ghh_am6}, similar to the flight model. A laboratory PC performed the phasementer back-end calculations (instead of the DMU onboard LTP) and produced the feedback signals for the test mass (3-axis PZT in our case). Because of the limited dynamic range of the used PZT, the criteria to switch between the acquisition modes had to be re-defined as follows: for a contrast value below 60\,\% the PZT performs a scan. The DC-servo is switched on when the contrast reaches 60\,\% and the DWS-servo takes over from 75\,\% until the end of the alignment. Implementation of the procedure with higher dynamic range PZTs is foreseen to study the convergence of the procedure under more stringent start parameters.

\begin{figure}[ht]
\includegraphics{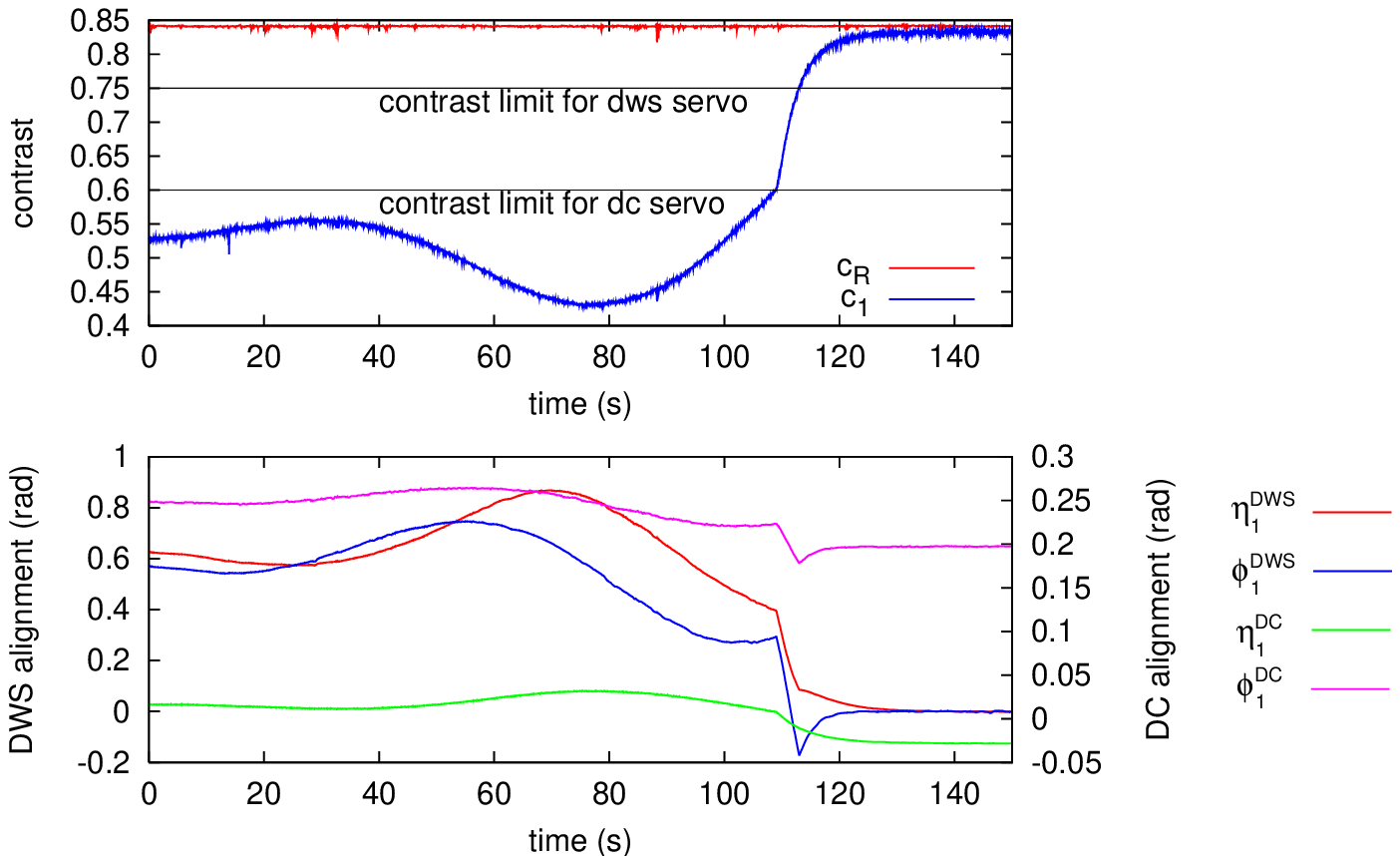}
\caption{\label{label1}Time evolution of the alignment: test mass is scanned until contrast in interferometer \textbf{1} (blue curve in upper graph) reaches the 60\,\% threshold. Then the DC-servo is switched on (signals green and magenta in the lower graph) and contrast value achieves 75\,\%. At this point the DWS-servo takes over until its error signals (red and blue in the lower graph) are zero and the contrast is optimal.}
\end{figure}
\begin{figure}[ht]
\begin{minipage}{18pc}
\includegraphics[width=18pc]{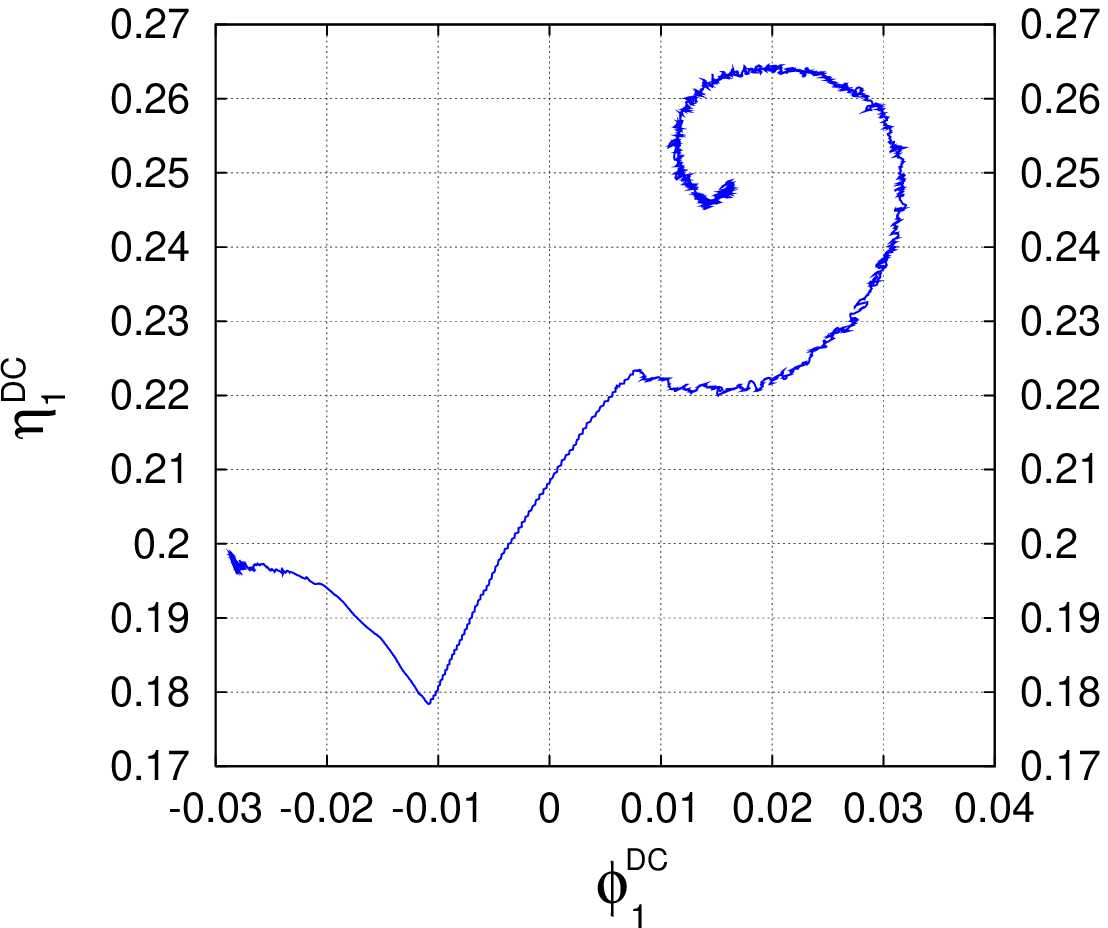}
\caption{\label{label2}Two-dimensional representation of the DC signals during the alignment procedure.}
\end{minipage}
\hspace{2pc}%
\begin{minipage}{18pc}
\includegraphics[width=18pc]{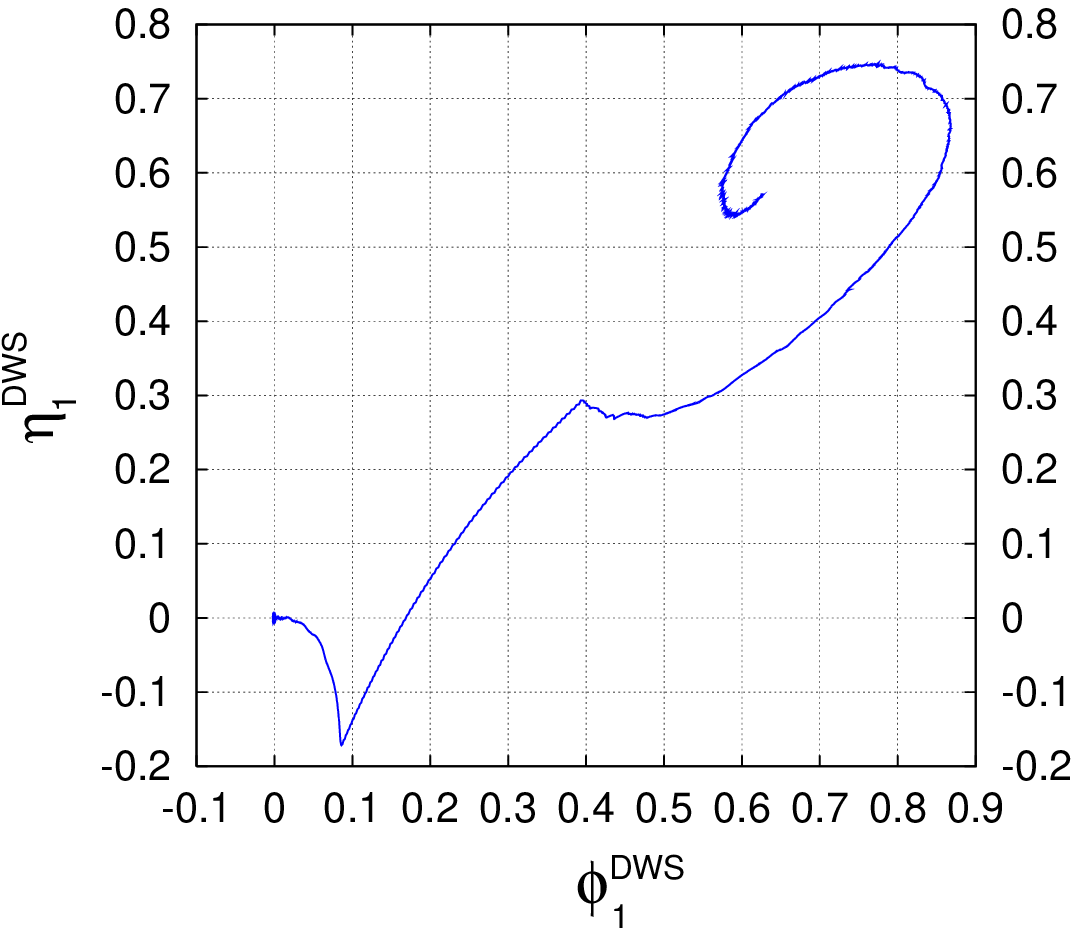}
\caption{\label{label3}Two-dimensional representation of the DWS signals, which are zero when the procedure is finished.}
\end{minipage} 
\end{figure}
Figures ~\ref{label1},~\ref{label2} and ~\ref{label3} show the successful implementation of the autonomous alignment.

\section{Diagnostic Procedures}
Further interferometric diagnostic procedures to be performed with LTP will deliver essential information about the on orbit environmental conditions for the LISA interferometry.

\subsection{Environmental phase noise}

The phase noise of single-mode fibers on orbit is important for LISA, as they are essential elements in the baseline architecture. 

On ground, this phase noise is dominated by environmental influences (such as thermal, seismic or pressure variations). On orbit, these fluctuations should be much smaller, but no reliable estimate of their magnitude and power spectrum is known.

\begin{figure}[ht]
\begin{minipage}{18pc}
\includegraphics[height=18pc,angle=-90]{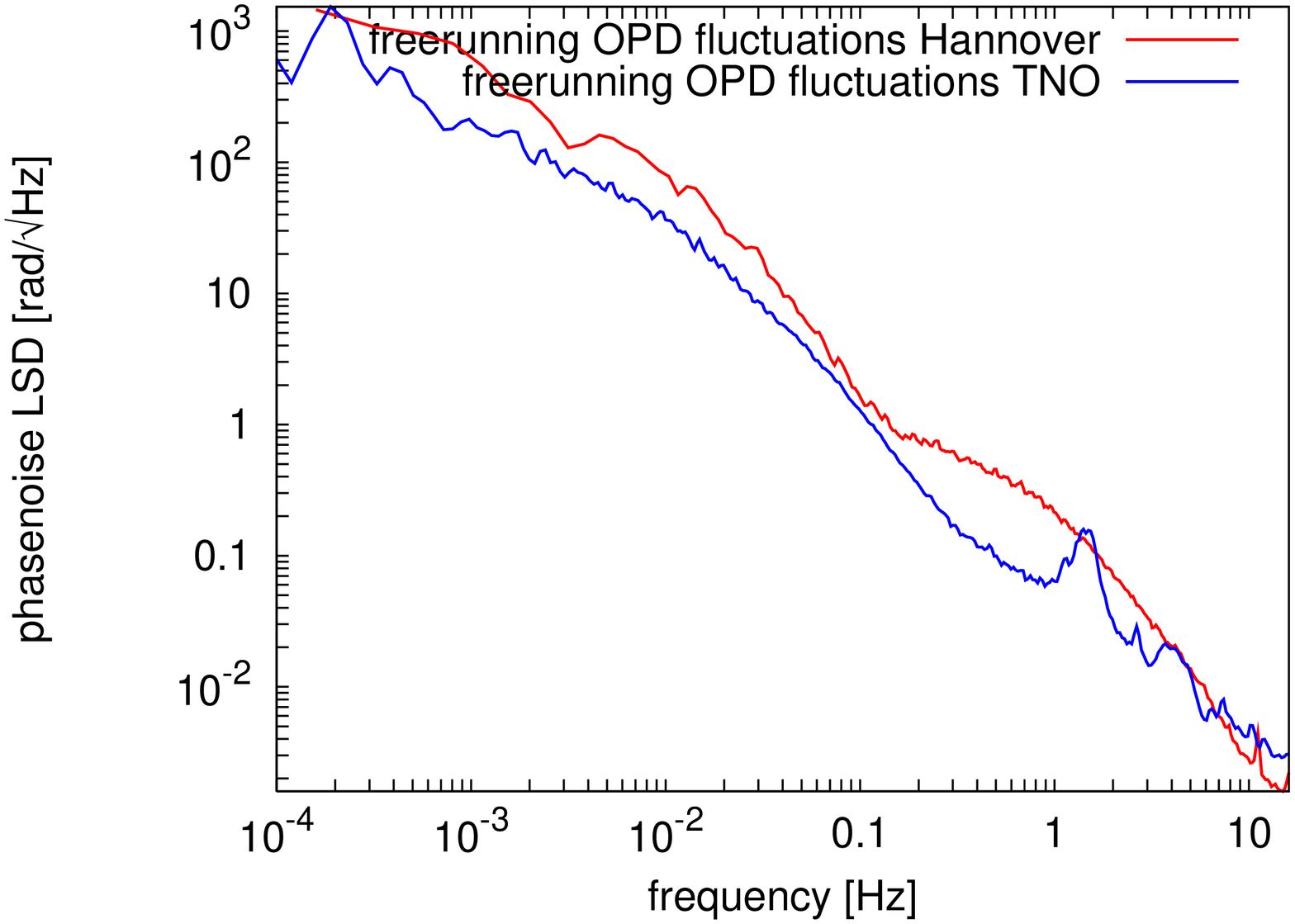}
\caption{\label{opd}OPD fluctuations measured at AEI-Hannover and TNO (The Netherlands).}
\end{minipage}\hspace{2pc}%
\begin{minipage}{18pc}
\includegraphics[height=18pc,angle=-90]{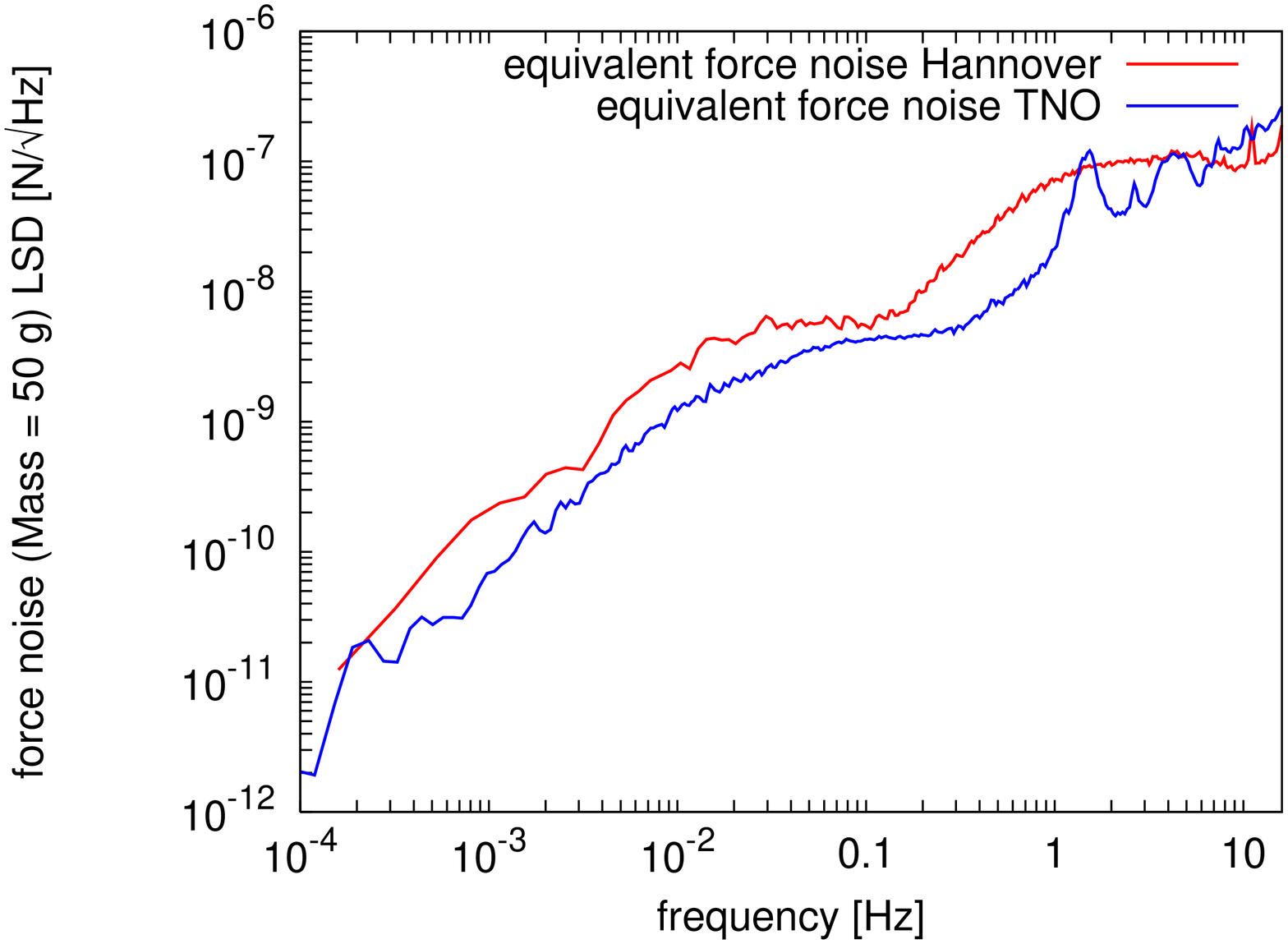}
\caption{\label{opd_force}Derived force fluctuations for a 50\,g OPD actuator.}
\end{minipage} 
\end{figure}

These optical pathlength difference fluctuations (OPD)~\cite{ghh_LISA,vwa_am6} appear as common mode noise in the output phase of the four LTP interferometers ($\varphi_{R}$,$\varphi_{1}$,$\varphi_{12}$,$\varphi_{F}$). Figure~\ref{opd} shows the linear spectral density (LSD) of $\varphi_{R}$ for two different laboratory conditions. 

Note that this common mode environmental noise cancels out in the measurement of test mass position fluctuations, as this is given by the substraction between $\varphi_{R}$ and $\varphi_{1}$ or $\varphi_{12}$, respectively~\cite{ghh_am4}. 

\subsection{Thermal influences}

\begin{figure}[ht]
\begin{minipage}{18pc}
\begin{center}
\includegraphics[width=14pc]{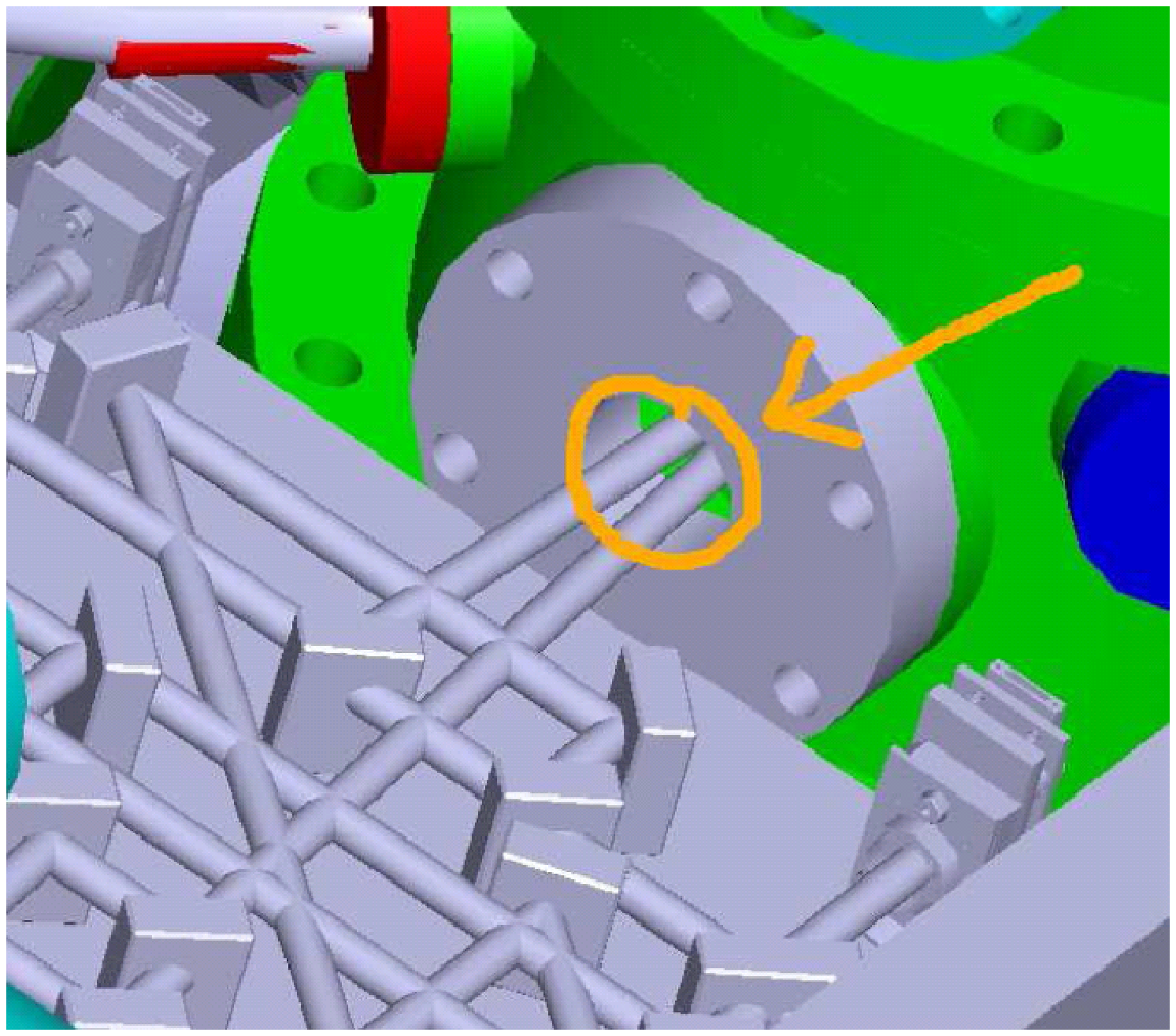}
\caption{\label{window}Optical window as interface between the optical bench and the vacuum enclosure containing the test mass}
\end{center}
\end{minipage}\hspace{2pc}%
\begin{minipage}{18pc}
\begin{center}
\includegraphics[width=18pc]{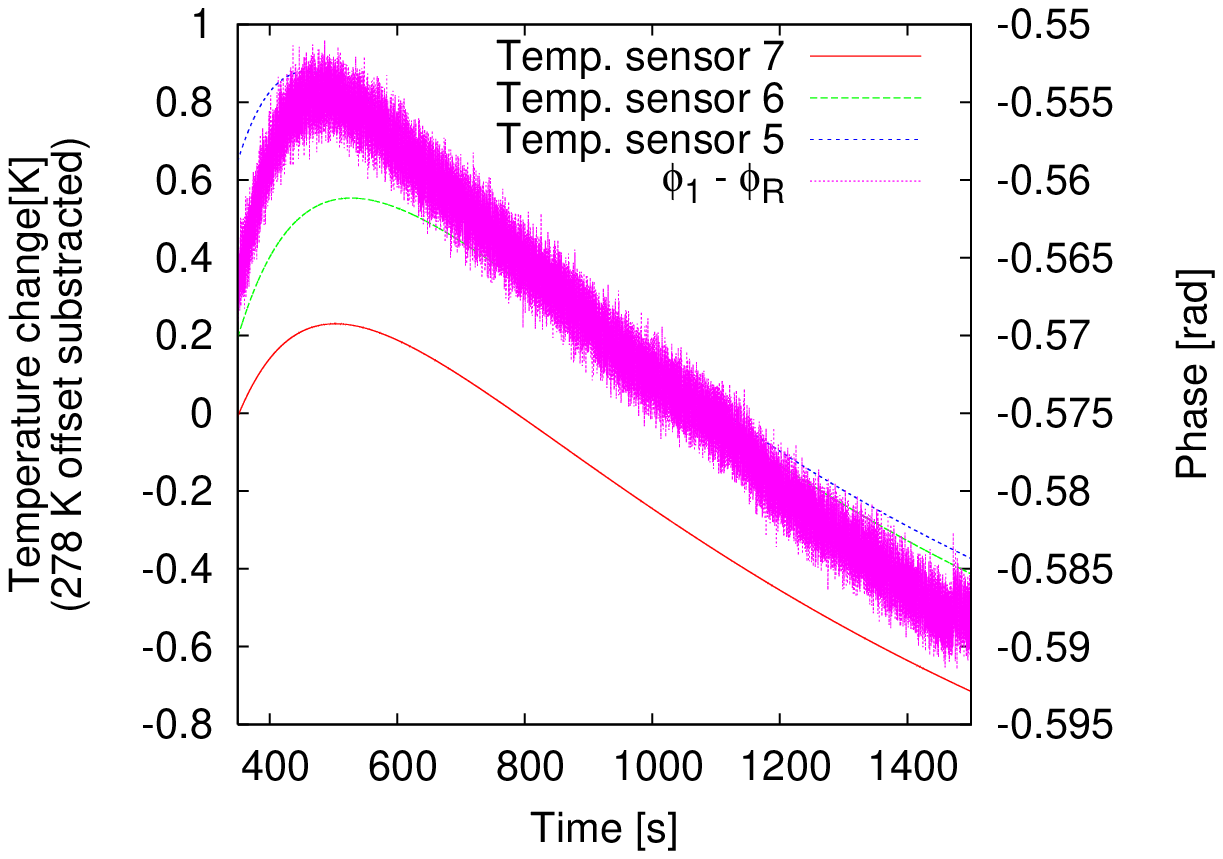}
\caption{\label{thermal}Pathlength evolution during a controlled temperature change in the optical window}
\end{center}
\end{minipage} 
\end{figure}

Thermal stability is expected to be an issue in LPF and even more in LISA. Several precision heaters and thermistors are foreseen in the LTP design for thermal analysis. In particular, the on-orbit thermo-optical properties of the optical windows that serve as interface between interferometer bench and test masses (see Figure~\ref{window}) are impossible to be accurately measured or predicted on ground. Hence, a controlled temperature change will be applied to several parts of the interferometer and the optical windows, while the resulting changes in pathlength and alignment are being monitored. 
Figure~\ref{thermal} shows a ground test of such measurements performed on unmounted prototypes of the window made of the same athermal glass as will be used in the flight model. 

The measurement of the pathlength fluctuations caused by temperature changes in the athermal glass can be used to determine the environmental temperature stability required to fullfill the aimed sensitivity of 10 pm/$\mathrm{\sqrt{Hz}}$ at 
1 mHz.

\subsection{Other procedures}
\begin{description}
\item[Absolute pathlength measurement:] The aim of this procedure is to obtain the difference between the expected test mass position and the real one during LTP operation. The LTP interferometer is designed so that when the test masses are in their nominal position, the length of the two arms of the \textbf{R}, \textbf{1} and \textbf{12} interferometer are equal and their output is insensitive to laser frequency noise~\cite{ghh_am4}. Applying laser frequency fluctuations (either noise or modulations) and measuring their effect in the interferometer output $\varphi_{1}$ or $\varphi_{12}$ allows the determination of the pathlength difference between the two interferometer arms and therefore the deviation of each test mass from its nominal position.

\item[Laser amplitude and frequency noise:] As an important input for LISA, the on orbit behaviour of the laser can be characterized during LTP operation. The amplitude injected in the optical bench will be monitored by means of the two photodiode labelled "Amp.stab." in Figure~\ref{layout}. The frequency noise of the laser shows up in the output of the frequency interferometer $\varphi_{\rm F}$~\cite{ghh_am4}. These signals will be used during conventional science runs to stabilize both amplitude and frequency of the laser. To obtain the on orbit unstabilized properties of the laser, a procedure has been defined in which the laser is left free-running and the mentioned signals are sent to ground. 

\item[Excess noise at picometer level~\cite{ghh_am6},\cite{vwa_am6},\cite{Robertson_LISA}:] 
A procedure has been defined to characterize this noise term.

\end{description}

\section{Conclusions}
As LTP enters its implementation phase, the on orbit operations of this experiment have to be defined. We have developed and tested several procedures emphasizing the alignment of the test masses with respect to the optical bench and its fully autonomous implementation.



\end{document}